\documentclass[
reprint,
superscriptaddress,
amsmath,amssymb,
aps,
pra,
10pt]{revtex4-2}

\usepackage{graphicx, subfigure}% Include figure files
\usepackage{dcolumn}% Align table columns on decimal point
\usepackage{bm}% bold math
\usepackage{hyperref}% add hypertext capabilities
\usepackage{physics}
\usepackage{dsfont}
\usepackage{amsmath}
\usepackage{braket}
\usepackage{amsfonts}
\usepackage{amssymb}
\usepackage{siunitx}
\usepackage[dvipsnames]{xcolor}
\usepackage{soul}
\sethlcolor{Yellow}

\begin{document}
	
	\preprint{APS/123-QED}
	\title{Lamb shift as a witness for quantum noninertial effects}
	\author{Navdeep Arya}
	\email{navdeeparya.me@gmail.com}
	\affiliation{Department of Physical Sciences, Indian Institute of Science Education \& Research (IISER) Mohali, Sector 81 SAS Nagar, Manauli PO 140306 Punjab India.}
	
	\author{Sandeep K.~Goyal}
	\email{skgoyal@iisermohali.ac.in}
	\affiliation{Department of Physical Sciences, Indian Institute of Science Education \& Research (IISER) Mohali, Sector 81 SAS Nagar, Manauli PO 140306 Punjab India.}
	
	\date{\today}
	
	\begin{abstract}
		    The sustained intense experimental activity around atomic spectroscopy and the resulting high-precision measurements of atomic spectral lines attracts interest in Lamb shift as a witness for noninertial effects in quantum systems.
			We investigate the Lamb shift in a two-level system, undergoing uniform circular motion, coupled to a quantum electromagnetic field inside a cavity. We show that when the separation between different cavity modes is large compared to the width of each cavity mode, both the inertial and purely-noninertial contributions to the Lamb shift are convergent. In addition, we find that the purely-noninertial Lamb shift maximizes away from the atomic resonance by an amount decided by the angular frequency of the circulating atom, lending itself to efficient enhancement by suitably tuning the cavity parameters. We argue that the purely-noninertial contribution becomes detectable at accelerations $\sim 10^{14}~\mathrm{m/s^2}$.
	\end{abstract}
	
	\maketitle

\section{Introduction}
In several studies, the noninertial motion has been shown to modify various properties of quantum systems, ranging from thermal signature in transition rates~\cite{unruh1976} and acceleration-induced transparency~\cite{Soda2022} to potentially detectable noninertial contributions to the geometric phase~\cite{Martinez2011,AryaPRD2022}. Moreover, interesting results have been obtained concerning the relationship between entanglement and noninertial motion~\cite{Alsing2003,*Alsing2006,Adesso2012,Bruschi2012,Friis2012,Bruschi2014,Restuccia2019,Toros2020, Toros2022}, for example, rotation can lead to the generation of entanglement~\cite{Toros2022}.
These studies extend our understanding of quantum physics beyond the well-understood domain of inertial reference frames into the domain of noninertial reference frames, thus laying the groundwork for ultimately investigating novel phenomena at the interface of quantum physics and gravity~\cite{Howl2018}.

Atoms, in such studies, are usually modeled as two-level systems coupling locally to a quantum field~\cite{unruh1976,Israel1979}. The response of such systems is controlled by the field correlation functions. The field correlators perceived by the atom are sensitive to its state of motion~\cite{fulling1973,unruh1976,davies1996,Letaw1980,Crispino2008} or the presence of gravity, leading to a distinct noninertial or gravitational~\cite{HawkingG1974,Hawking1975} contribution to the atomic response. The resulting effects are usually very weak and require extreme acceleration or gravitational field for a detectable signature. 

Different studies~\cite{Rogers1988,Tajima1999,fuentes2010,scully2003,Jaffino2022,Vanzella2001,Barshay1978,*Barshay1980,*Kharzeev2006,bell1983,*bell1987,*unruh1998,Dolan2020,Vriend2021,Lynch2021,Matos2021} have investigated various properties of noninertial quantum systems under varied conditions, seeking appreciable noninertial signatures and ease of measurement in laboratory settings. An observable of interest in this context is the Lamb shift, or the radiative energy shift in general (though in this work we will use the two terms interchangeably). 

In this work, we are interested in the correction to the radiative energy shifts originating from the atom's noninertial motion. The total radiative energy shift in an atom on a noninertial trajectory has two contributions: inertial and \textit{purely}-noninertial. The additional purely-noninertial contribution comes due to the acceleration of the atom. The atom is assumed to be coupled to a quantum electromagnetic (EM) field inside a cavity. In particular, we are interested in the effect of the modified density of field states inside the cavity on the inertial and purely-noninertial contributions to the Lamb shift.

The Lamb shift is a shift in the energy levels of an atom due to the atomic electron's coupling to a quantum electromagnetic field~\cite{Lamb1947,Bethe1947}. The Lamb shift in inertial atoms has been measured with great precision using different experimental methods~\cite{Hagley1994,Weitz1994,Berkeland1995,Bezginov2019}. As already mentioned, the response of atoms coupled to a quantum field is determined by the field correlators, which depend on the atom's trajectory and therefore lead to a noninertial signature in the radiative energy shifts. The energy shift can be an observable of interest for the detection of the effects of acceleration~\cite{Audretsch1995PRA,Audretsch1995IOP,Passante1998} and gravity~\cite{Parker1980,Fischbach1981,Parker1982,Zhou2010,Zhou2012} owing to the theoretical and experimental advances that atomic spectroscopy has made~\cite{CODATA2018,Beyer2016,Beyer2017,Fleurbaey2018,Bezginov2019,Grinin2020}. Particularly, transitions in Hydrogen atom have been measured with precision in the $10^{-11} - 10^{-12}$ range for optical transitions and $10^{-5} - 10^{-6}$ range for microwave transitions~\cite{CODATA2018,Karr2020}.  
Such precise measurements of transition frequencies enable the determination of various corrections to the spectral lines, including those coming from the atomic electron's interaction with a quantum electromagnetic field. 

The Lamb shift in atomic systems is predominantly a nonrelativistic phenomenon in the sense that a major contribution to the Lamb shift comes from the atomic electron's coupling to the field modes with energy less than the electron's rest mass energy~\cite{Milonni1994,sakurai2006adv}. Therefore, we will focus on such a nonrelativistic treatment. The Lamb shift in a two-level atom on a stationary worldline~\cite{Letaw1981}, coupled to a quantum scalar field in free space, has a logarithmically divergent inertial contribution and a finite correction coming due to the atom's noninertial motion~\cite{Audretsch1995PRA,Audretsch1995IOP}. The logarithmically divergent inertial contribution necessitates the introduction of an ultraviolet cutoff~\cite{sakurai2006adv}. 
Such cutoffs employed in the nonrelativistic calculations of the radiative corrections to matter properties lead to cutoff-sensitive results. See Ref.~\cite{Milonni1994-a} for a typical example related to the calculation of anomalous magnetic moment of an electron.

The quantum electrodynamical properties of an atom interacting with an EM field get modified if the density of field states is changed, for example, by introducing conducting mirrors~\cite{Barton1970,*Barton1979,*Barton1987}. The EM cavities have been fruitfully studied in the context of amplification and isolation of the noninertial quantum field theoretic effects~\cite{lochan2020,AryaPRD2022,Jaffino2022}. The radiative energy shift, in particular, depends sensitively on the density of field modes~\cite{Belov1989}.
Using an EM cavity, the transition rates of an atom can be either amplified or inhibited~\cite{Purcell1995}. Therefore, as the Lamb shift arises due to the absorption and emission of virtual photons by the atomic electron~\cite{API_Tannoudji}, better control over the energy shifts can be obtained using an EM cavity~\cite{Walther2006}. The radiative energy shifts in an inertial atom in the presence of a mirror~\cite{Wilson2003}, in an atom placed between parallel metal plates~\cite{Marrocco1998}, and in an atom placed inside a confocal resonator~\cite{Heinzen1987} have been measured in various experiments which, if the purely-noninertial contribution is appreciable, can conceivably be extended to atoms on noninertial trajectories.

Here, we study the Lamb shift in an atom undergoing uniform circular acceleration and coupled to an electromagnetic field inside a cavity. Specifically, we focus on the behavior of the inertial and noninertial contributions to the total Lamb shift as a function of the cavity's normal frequency and discuss the detectability of the noninertial contribution. We show that when the separation between different cavity modes is large as compared to the width of each cavity mode, both the inertial and purely-noninertial contributions to the Lamb shift are convergent, leading to cutoff-independent results. In addition, we find that the purely-noninertial Lamb shift maximizes away from the atomic resonance by an amount decided by the angular frequency of the circulating atom, lending itself to efficient enhancement by suitably tuning the cavity parameters. We argue that the purely-noninertial contribution can be detected at accelerations $\sim 10^{14}~\mathrm{m/s^2}$.

This paper is organized as follows. In Sec.~\ref{sec:background} we discuss the Lamb shift in an open quantum system from the perspective of the Lindblad master equation formalism. In Sec.~\ref{Sec: Atom-Cavity Setup}, we detail the atom-cavity setup employed in this work. In Sec.~\ref{sec:results}, we use the Lindblad master equation formalism to obtain the Lamb shift in a two-level system on a circular trajectory inside an EM cavity. Finally, in Sec.~\ref{sec:discussion} we discuss the results and conclude with a discussion on the outlook for the study carried out in this paper.

\section{Background : Lamb Shift}\label{sec:background}
In this section, we discuss the Lamb shift in a small system $S$ due to its coupling to a large reservoir $B$.
The intrinsic dynamics of $S$ and $B$ are governed by the Hamiltonians $H_S$ and $H_B$, respectively. The Schrodinger picture interaction Hamiltonian between the system and the reservoir can be written as
\begin{equation}
	H_I = \sum_{\alpha} A_{\alpha} \otimes \tilde{B}_{\alpha},
\end{equation}
where $A_{\alpha} = A^{\dagger}_{\alpha}$ and $\tilde{B}_{\alpha} = \tilde{B}^{\dagger}_{\alpha}$ are the system and the reservoir operators, respectively.
The Lindblad (interaction picture) master equation governing the dynamics of the system $S$ is given by~\cite{Breuer2007}
\begin{equation}\label{MasterEq}
	\derivative{\rho_S(\tau)}{\tau} = -\frac{i}{\hbar} \comm{H_{\text{LS}}}{\rho_S(\tau)} + \mathcal{D}\left(\rho_S(\tau)\right),
\end{equation}
where $\tau$ is the proper time, $\rho_S(\tau)$ is the system's density operator and $\mathcal{D}(\rho_S(\tau))$ is called the dissipator of the master equation as it controls the dissipation and decoherence in the system. $H_{\text{LS}}$ is known as the Lamb shift Hamiltonian as it leads to a renormalization of the unperturbed energy levels induced by the system-reservoir coupling.

The Lamb shift Hamiltonian is given by
\begin{equation}\label{HLS}
	H_{\text{LS}} = \sum_{\nu} \sum_{\alpha,\beta} \hbar S_{\alpha \beta}(\nu) A_{\alpha}^{\dagger}(\nu) A_{\beta}(\nu),
\end{equation}
where
\begin{equation}
	S_{\alpha \beta}(\nu) \equiv  \frac{1}{2 \pi} \int_{-\infty}^{\infty} \dd{\lambda} \mathcal{G}_{\alpha \beta}(\lambda) \text{P.V.} \left(\frac{1}{\nu - \lambda}\right),
\end{equation}
with 
\begin{equation}
	\mathcal{G}_{\alpha \beta}(\lambda) \equiv \frac{1}{\hbar^2} \int_{-\infty}^{\infty} \dd{\tau_-} e^{i \lambda \tau_-} \tr_{\text{B}}\left( \tilde{B}_{\alpha}(\tau_2) \tilde{B}_{\beta}(\tau_1)\right),
\end{equation}
$\tau_- \equiv \tau_2 - \tau_1$, P.V. denoting the Cauchy Principal value integral, and $\tr_{\text{B}}(\cdot)$ denoting the trace over the reservoir degrees of freedom. Note that $\mathcal{G}_{\alpha \beta}(\lambda)$ is the Fourier transform of the two-point reservoir correlation function.
Further, the $\sum_{\nu}$, where $\nu \equiv \varepsilon' - \varepsilon$, is extended over all eigenvalues $\varepsilon$ and $\varepsilon'$ of $H_{\text{S}}$ with a fixed energy difference $\nu$. The $ A_{\beta}(\nu)$ are the eigenoperators of the system Hamiltonian $H_S$ and are defined as $A_{\alpha}(\nu) \equiv \sum_{\nu} \Pi(\varepsilon) A_{\alpha} \Pi(\varepsilon') $, where $\Pi(\varepsilon)$ is the projector on the eigenspace belonging to the eigenvalue $\varepsilon$~\cite{Lidar2019notes,Breuer2007}. The $A_{\alpha}(\nu)$ are also known as the Lindblad operators.

Now, consider $S$ to be a two-level atom with the excited state $\ket{e}$ and the ground state $\ket{g}$, interacting with a quantum electromagnetic field inside an electromagnetic cavity. The proper frequency gap between the two atomic levels is $\Omega_0$ and the atom carries an electric dipole moment four-vector $\hat{d}'^{\mu} = (\hat{d}'^{0},\hat{\vb{d}}')$. In the interaction picture, the dipole moment operator $\hat{\vb{d}}'(\tau)$ is given by $\hat{\vb{d}}'(\tau) = \vb{d}' \sigma_{-} \exp(-i \Omega_0 \tau) + \vb{d}'^* \sigma_{+} \exp(i \Omega_0 \tau) $, where $\vb{d}' \equiv \bra{g} \hat{\vb{d}}'(\tau = 0) \ket{e}$, and $\sigma_{+} = \sigma^{\dagger}_{-} = \ket{e} \bra{g}$ is the step-up operator for the atomic states.
The Lindblad operators for $S$ are given by~\cite{Breuer2007}
\begin{equation}\label{LindbladOps}
	\vb{A}(\Omega_0) = \vb{d}' \sigma_-, ~ \vb{A}(-\Omega_0) = \vb{d}'^* \sigma_+. 
\end{equation} 
The electromagnetic field is assumed to be in the inertial vacuum state $\ket{0}$. The interaction Hamiltonian between the atom and the electromagnetic field is given by $H_I = - \hat{d}^{\mu}E_{\mu}$~\cite{Anandan2000}, where $E_{\mu} \equiv F_{\mu \nu} u^{\nu}$, $F_{\mu\nu}$ is the electromagnetic field strength tensor and $u^{\nu}$ is the four-velocity of the atom. The interaction Hamiltonian takes the form $H_I = - \hat{\vb{d}}' \cdot \vb{E'}$ in the rest frame of the atom, where $\vb{E}'$ is the electric field 3-vector as seen by the atom. Throughout this paper, primed quantities correspond to the atom's rest frame.

From Eqs.~\eqref{HLS} and \eqref{LindbladOps}, for the two-level system we have
\begin{multline}
		H_{\text{LS}} = \hbar \sum_{\alpha,\beta} \Big( S_{\alpha \beta}(\Omega_0) (d'_{\alpha} \sigma_-)^{\dagger} d'_{\beta} \sigma_- \\
		+ S_{\alpha \beta}(-\Omega_0) (d'^*_{\alpha} \sigma_+)^{\dagger} d'^*_{\beta} \sigma_+ \Big).
\end{multline}
For simplicity, we assume that $\vb{d}' = (0, d', 0)$, and obtain
\begin{equation}
		H_{\text{LS}} = \hbar \abs{\vb{d}'}^2 \left( S_{22}(\Omega_0) \sigma_+ \sigma_- +  S_{22}(-\Omega_0)  \sigma_- \sigma_+ \right).
\end{equation}
As is clear from the form of $H_{\mathrm{LS}}$ in above equation, it induces transitions between the two atomic levels mediated by virtual photons.
The Lamb shift, $\Delta$, is obtained as
\begin{equation}
	\begin{split}
		\hbar\Delta &\equiv \ev{H_{\text{LS}}}{e} - \ev{H_{\text{LS}}}{g}\\
		&= \hbar\abs{\vb{d}'}^2 \left(S_{22}(\Omega_0) - S_{22}(-\Omega_0)\right),
	\end{split}
\end{equation}
that is,
\begin{equation}\label{LS1}
	\Delta = \frac{\abs{\vb{d}'}^2}{2 \pi} \int_{-\infty}^{\infty} \dd{\nu'} \mathcal{G}'_{22}(\nu') \text{P.V.} \left(\frac{1}{\nu' + \Omega_0} - \frac{1}{\nu' - \Omega_0}\right),
\end{equation}
where
\begin{equation}\label{FT}
	\mathcal{G}'_{22}(\nu') = \frac{1}{\hbar^2} \int_{-\infty}^{\infty} \dd{\tau_-} e^{i \nu'  \tau_-} G'^+_{22}(\tau_-),
\end{equation}
with $G'^+_{22}(\tau_-) \equiv \ev{E'_y(\tau_2)E'_y(\tau_1)}{0}$ being the positive frequency vacuum Wightman function.

%%%%%%%%%%%%%%%%%%%%%%%%%%%%%%%%%%%%%%%%%%%%%%%%%%%%%%%%%%%%%%%%%%%%%%%%%%%%%%%%%%%%%%%%%%%%%%%%%%%%%%%%%%%%%%%%
\section{Atom-cavity setup}\label{Sec: Atom-Cavity Setup}
In this section, we discuss the atom-cavity setup employed in this study. The atomic transition rates depend on the field spectral density in the resonant mode, that is, in the mode with frequency $\Omega_0$. From Eq.~\eqref{LS1}, however, note that since P.V.$\left(1/(\nu \pm \Omega_0)\right)$ vanishes for $\nu = \pm \Omega_0$ and behaves as $1/(\nu \pm \Omega_0)$ away from $\nu = \pm \Omega_0$, the radiative energy shift depends on the field spectral density in all the modes, except the resonant mode, with a weight that falls off away from the resonant mode~\cite{heitler1954}. 
If we consider a cavity such that the frequency separation between different cavity modes is large as compared to the width of each cavity mode, then due to the presence of the factor P.V.$\left(1/(\nu \pm \Omega_0)\right)$ in Eq.~\eqref{LS1} the dominant contribution to the energy shift will come from the modes in the vicinity of the cavity's normal frequency $\omega_{\rm c}$, as compared to the contribution of the higher frequencies supported by the cavity at $n\omega_{\rm c}, n > 1$. We assume that the cavity modes are separated in frequency by much more than their width (i.e., a cavity with a high quality-factor).
If the inertial rate at which the radiation deposited in the cavity is damped is $\omega_c/Q$, where $Q$ is the quality factor of the cavity, then the density of field states inside the cavity is given by~\cite{Scully_Zubairy,*Lewenstein1993,Heinzen1987}
\begin{equation}\label{dos}
	\rho(\omega_k) = \frac{1}{\pi} \frac{(\omega_c/Q)}{(\omega_c/Q)^2 + (\omega_k - \omega_c)^2}.
\end{equation}

\begin{figure*}
	\centering
	\subfigure[]{
		\includegraphics[width=0.4\linewidth]{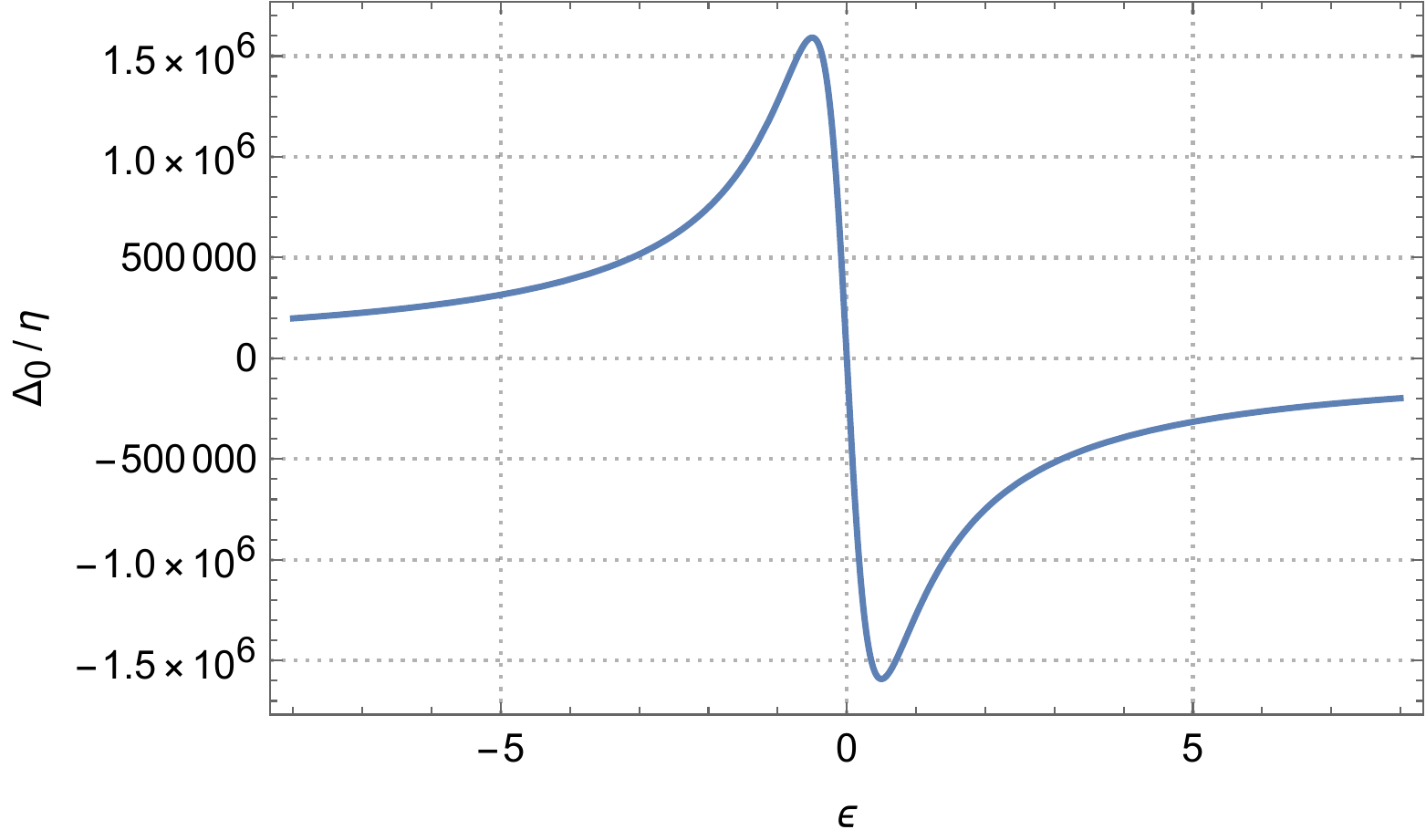}
		\label{fig:LS(rest)}}
	\subfigure[]{
		\includegraphics[width=0.4\linewidth]{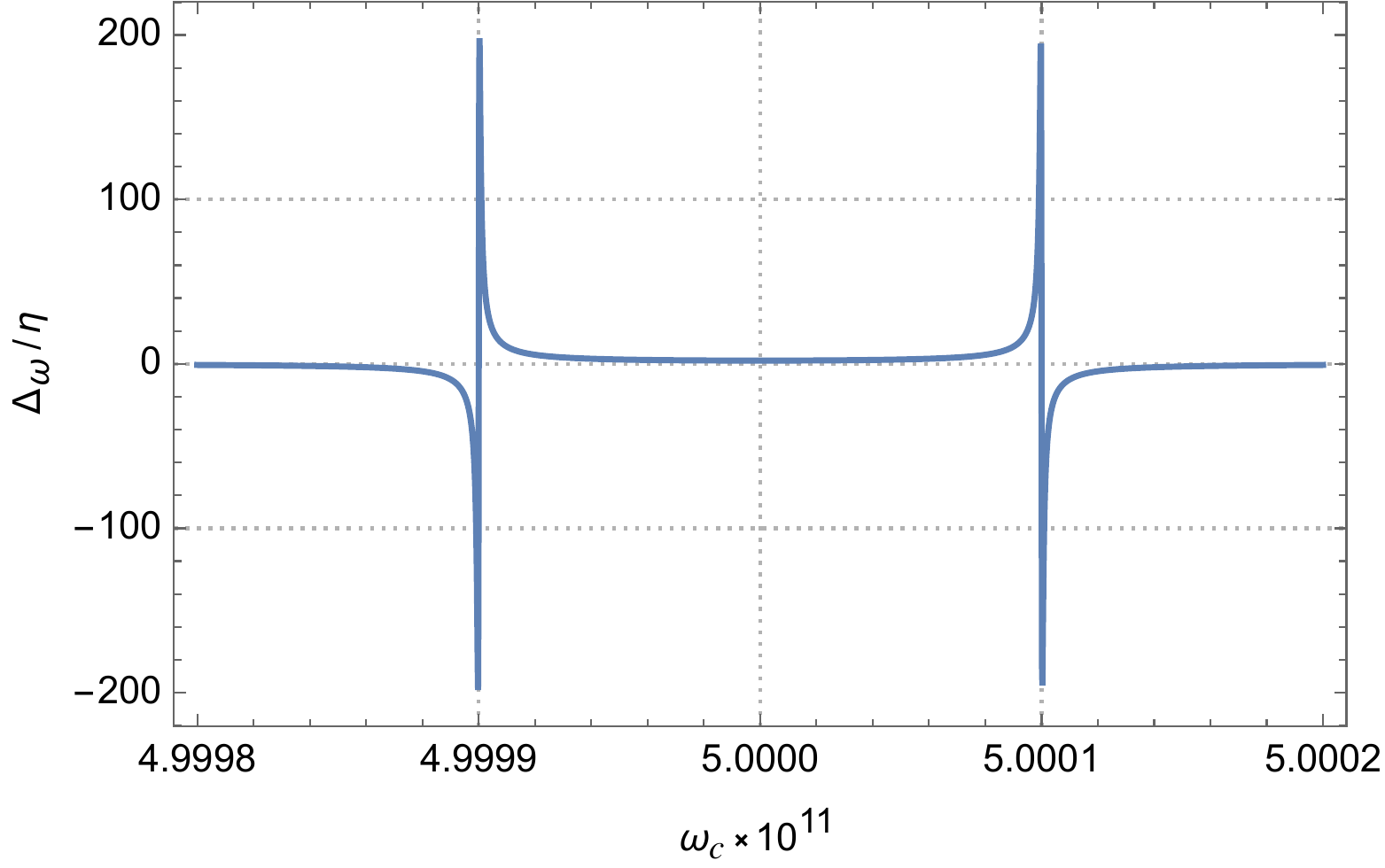}
		\label{fig:LS(5x10^11)}}
	\caption{Fig.~\eqref{fig:LS(rest)} plots $\Delta_0/\eta$ as a function of the cavity's normal frequency [See Eq.~\eqref{LSrest}], where we have written $\omega_{c} = 10^7 + \epsilon$. We see that the Lamb shift in an inertial atom maximizes when the cavity is tuned to a frequency in the vicinity of the atomic resonance and falls off on both sides of this region. Since we are considering a two-level system with $\Omega_0 = 10^7~\mathrm{Hz}$, we take $\omega_c \sim 10^{7}~\mathrm{Hz}$. For a cavity with $Q \sim 10^7$, we take $V \sim 10^{-5}~\mathrm{m^3}$ to ensure consistency with Eq.~\eqref{badcavity}. Therefore $\eta \sim 10^{-9}~\mathrm{Hz}$ ($\eta \equiv \abs{\vb{d}' }^2/3 \pi \hbar \epsilon_0 V$), which leads to $\Delta_0 \sim 10^{-3}~\mathrm{Hz}$. For an atom on a circular orbit with angular velocity $\omega$,  however, there are additional peaks in the neighborhood of $\omega_c = \omega \pm \bar{\Omega}_0$. See Fig.~\ref{fig:LS&Ratio} as well. Fig.~\eqref{fig:LS(5x10^11)} gives a plot of $\Delta_{\omega}/\eta$, defined in Eq.~\eqref{NonInLS}, as a function of the cavity's normal frequency for $\omega = 5 \times 10^{11}~\mathrm{Hz}$. In this case, $\omega_c \sim \omega \pm \Omega_0$ and $\omega \gg \Omega_0$. Therefore, for $Q \sim 10^7$, we take $V \sim 10^{-9}~\mathrm{m^3}$, and $R = 10^{-5}~\mathrm{m}$. This corresponds to $\eta \sim 10^{-5}~\mathrm{Hz}$, and an average acceleration $a = \omega^2 R \sim 10^{18}~\mathrm{m/s^2}$ and leads to a purely-noninertial Lamb shift $\Delta_{\omega} \sim 10^{-3}~\mathrm{Hz}$. Such accelerations can be achieved with electrons in storage rings~\cite{bell1983}.}
	\label{fig:LS}
\end{figure*}

Further, to ensure the validity of the Markovian approximation~\cite{Breuer2007}, which is inherent in the derivation of the master equation~\eqref{MasterEq}, we will work in the bad-cavity regime, that is,
\begin{equation}\label{badcavity}
	g \ll \kappa,
\end{equation} 
where, for a cavity of volume $V$, $g \equiv \abs{\vb{d}'} \sqrt{\omega_{c}/(2 \hbar \epsilon_0 V)}$ is the atom-cavity field coupling constant and $\kappa \equiv \omega_c/Q$ is the cavity-field decay rate~\cite{ReisererRMP2015}. For a given atom and a given normal frequency $\omega_c$ of the cavity, the condition $g \ll \kappa$ will decide the allowed values of $V$ and $Q$ consistent with the Markovian approximation. See the caption of Fig.~\eqref{fig:LS}.

The variation of P.V.$\left(1/(\nu \pm \Omega_0)\right)$ and the density of field modes $\rho(\nu)$, entering Eq.~\eqref{LS1} through the field correlation function, decide the magnitude of the radiative energy shift.
%%%%%%%%%%%%%%%%%%%%%%%%%%%%%%%%%%%%%%%%%%%%%%%%%%%%%%%%%%%%%%%%%%%%%%%%%%%%%%%%%%%%%%%%%%%%%%%%%
\section{Results}\label{sec:results}
We now consider a two-level atom on a circular trajectory of angular frequency $\omega$ and radius $R$ inside an electromagnetic cavity of volume $V$. 

In the following subsection, we determine the Fourier transform of the two-point vacuum Wightman function as given in Eq.~\eqref{FT} and then use Eq.~\eqref{LS1} to obtain the Lamb shift in the subsequent subsection.
\subsection{Fourier transform of the field correlation function}
We consider an atom on a circular trajectory of radius $R$ and angular frequency $\omega$ such that its position four-vector in the lab frame is given by
\begin{equation}\label{eq:atomic_four_vec}
	\begin{split}
		&x^{\mu}(\tau) = (t(\tau), x(\tau), y(\tau), z(\tau))\\
		&= \left(c\gamma \tau, x_0 + R \cos(\omega \gamma \tau), 0, z_0 + R \sin(\omega \gamma \tau)\right),
	\end{split}
\end{equation}
where $\gamma \equiv (1-\zeta(\omega))^{-1/2}$, and $\zeta(\omega) \equiv \omega^2 R^2/c^2$.
To compute $\mathcal{G}'_{22}(\nu)$, we need the positive-frequency vacuum Wightman function $G'^{+}_{22}(\tau_-)$ in the atom's frame. We start by noting that $G'^{+}_{\mu\nu} \equiv \ev{E'_{\mu}(\tau_2)E'_{\nu}(\tau_1)}{0}$ can be obtained from its counterpart $G^{+}_{\mu\nu} \equiv \ev{E_{\mu}(x^{\lambda}_2)E_{\nu}(x^{\lambda}_1)}{0}$ in the lab frame using the tensor transformation between the two frames. Consequently, we have
\begin{equation}\label{wightman}
	G'^{+}_{22} = \sum_{\alpha \beta} \pdv{x^{\alpha}}{x'^{2}} \pdv{x^{\beta}}{x'^{2}} G^{+}_{\alpha \beta} = G^{+}_{22},
\end{equation}
where we have used the fact that the rotating $(\tau, x', y', z')$ and the inertial coordinates ($t, x, y, z$) are related as 
\begin{align}
	x' & = x - x_0 - R \cos\omega t, \nonumber \\ \noalign{\vskip5pt}
	y' & = y,\nonumber  \\ \noalign{\vskip5pt}
	z' & = z - z_0 - R \sin\omega t, \nonumber  \\ \noalign{\vskip5pt}
	\tau &= \left(1-\omega^2 R^2/c^2\right)^{1/2} t, \label{eq:coordinates}
\end{align} 
with $(x_0, 0, z_0)$ being the center of the circular trajectory.

The electric field perceived by the atom and as reported by the inertial observer is given by $E_{\mu}(x^{\lambda}) = F_{\mu \nu}(x^{\lambda}) u^{\nu}$, where $u^{\nu} \equiv \dv*{x^{\nu}(\tau)}{\tau}$ is the atomic four-velocity in the lab frame. Therefore, using Eq.~\eqref{eq:atomic_four_vec} we have
\begin{equation}\label{E_2}
	\begin{split}
		E_2 =\gamma \left[ E_y - \omega R \left\{ B_z \sin(\omega \gamma \tau) - B_x \cos(\omega \gamma \tau) \right\} \right],
	\end{split}
\end{equation}
which we will use for the computation of $G^{+}_{22}$ below.
The calculation of $\mathcal{G}'_{22}(\nu)$ can be further simplified by noting that under the Markovian approximation, we have the identity~\cite{Lidar2019notes}
\begin{multline}\label{LidarInt}
		\int_{-\infty}^{\infty} \dd{u} e^{i \omega u} \expval{E_{j}(u) E_{k}(0)}{0} \\
		= \lim_{T \rightarrow \infty} \frac{1}{T} \int_{0}^{T} \dd{\tau_2} \int_{0}^{T} \dd{\tau_1} e^{i \omega (\tau_2 - \tau_1)} \expval{E_{j}(\tau_2) E_{k}(\tau_1)}{0},
\end{multline}
where $j, k = 1,2,3$.
Using Eq.~\eqref{LidarInt} in the atom's comoving frame we can write
\begin{equation}
	\begin{split}
		&\mathcal{G}'_{22}(\nu) = \frac{1}{\hbar^2} \int_{-\infty}^{\infty} \dd{\tau_-} e^{i \nu \tau_-} G'^+_{22}(\tau_-) \\
		&= \lim_{T \rightarrow \infty} \frac{1}{\hbar^2 T} \int_{0}^{T} \dd{\tau_2} \int_{0}^{T} \dd{\tau_1} e^{i \nu (\tau_2 - \tau_1)} \expval{E'_{2}(\tau_2) E'_{2}(\tau_1)}{0},
	\end{split}
\end{equation}
and further using Eq.~\eqref{wightman} and $\dd{\tau} = \dd{t}/\gamma$, we have
\begin{equation}\label{CalG2b}
	\begin{split}
		\mathcal{G}'_{22}(\nu) &= \lim_{T \rightarrow \infty} \frac{1}{\gamma^2 \hbar^2 T} \int_{0}^{\gamma T} \dd{t_2} \int_{0}^{\gamma T} \dd{t_1} e^{i \bar{\nu} (t_2 - t_1)} \\
		& \times \expval{E_{2}(x^{\mu}_2) E_{2}(x^{\mu}_1)}{0},
	\end{split}
\end{equation}
where $\bar{\nu} \equiv \nu/\gamma$.

As shown in Appendix~\ref{ap:crossterms}, form Eq.~\eqref{CalG2b} we can obtain
\begin{multline}\label{CalG2a}
		\mathcal{G}'_{22}(\bar{\nu}) = \frac{\gamma}{\hbar^2} \int_{-\infty}^{\infty} \dd{t_-} e^{i \bar{\nu} t_-} \Big[\bra{0}E^{y}(x^{\mu}_2) E^{y}(x^{\mu}_1) \ket{0} \\
		+ \frac{R^2\omega^2}{2} \cos(\omega t_-)\bra{0}B^{z}(x^{\mu}_2) B^{z}(x^{\mu}_1) \ket{0} \\
		+ \frac{R^2\omega^2}{2} \cos(\omega t_-) \bra{0}B^{x}(x^{\mu}_2) B^{x}(x^{\mu}_1) \ket{0}  \Big],
\end{multline}
where we have also explicitly indicated the dependence of $\mathcal{G}'_{22}$ on $\bar{\nu}$.

The inertial frame electric and magnetic field operators in Coulomb gauge inside a quantization volume $V$ are given by~\cite{Gerry2004}
\begin{subequations}\label{E&B}
	\begin{equation}\label{E}
		\vb{E}[x(\tau)] = i \sum_{\vb{k},\lambda} \mathcal{E}_k \epsilon_{\vb{k},\lambda} \left(a_{\vb{k},\lambda} e^{-i\left(\omega_k t(\tau) - \vb{k}.\vb{x}(\tau)\right)}- \text{h.c.}\right),
	\end{equation}
	\begin{equation}\label{B}
		\vb{B}[x(\tau)] = \frac{i}{c} \sum_{\vb{k}\lambda} \mathcal{E}_k \left(\vu{k} \times \epsilon_{\vb{k}\lambda}\right) \left(a_{\vb{k}\lambda} e^{-i(\omega_k t(\tau) - \vb{k \cdot x(\tau)})} - \text{h.c.}\right),
	\end{equation}
\end{subequations}
respectively, where $\mathcal{E}_k \equiv \sqrt{\hbar \omega_k/(2\epsilon_0 V)}$; $\epsilon_{\vb{k},\lambda}$ with $\lambda = 1,2$ are the two orthogonal polarization vectors; $[a_{\vb{k}\lambda},a^{\dagger}_{\vb{k}'\lambda'}]= \delta_{\vb{kk'}} \delta_{\lambda \lambda'}$; and h.c. denotes Hermitian conjugate.
 
Using Eqs.~\eqref{E&B} we can evaluate the different field correlators~\cite{lochan2020} appearing in Eq.~\eqref{CalG2a} and, to first order in $\zeta(\omega)$, obtain
\begin{multline}\label{CalG3}
	\mathcal{G}'_{22}(\bar{\nu}) = \frac{\gamma}{3\pi\hbar \epsilon_0 V} \int_{0}^{\infty} \dd{\omega_k}\rho(\omega_k)  \omega_k \Big[\delta( \bar{\nu} - \omega_k) \\
	+\frac{R^2 \omega^2}{2c^2}  \frac{ 1}{2} \left[\delta( \bar{\nu}  - \omega_k + \omega) + \delta(\bar{\nu}  - \omega_k - \omega)\right] - \frac{2\omega_k^2 R^2}{5c^2} \times \\
	\Big\{ \delta( \bar{\nu} - \omega_k)  - \frac{1}{2}  \left(\delta( \bar{\nu} + \omega -\omega_k) + \delta( \bar{\nu} - \omega -\omega_k)\right) \Big\} \Big],
\end{multline}
where we have used
\begin{equation}
	\sum_{\vb{k}} \to \int \dd{\Omega_{\vb{k}}}\int \frac{\dd{k}\rho(k)}{(2\pi)^3},
\end{equation}
and $\rho(k) \dd{k} = \rho(\omega_k) \dd{\omega_k}$, with $\rho(\omega_k)$ being the density of field modes inside the cavity as given in Eq.~\eqref{dos}.
 Also note that to obtain Eq.~\eqref{CalG3} we have assumed a closed cavity. If we had assumed, for example, a concentric resonator instead of a closed cavity, the angular integrals in the field correlators appearing in Eq.~\eqref{CalG2a} would have split into two parts, one with the free space field mode density $\rho_{\text{free}}(\omega_k) = \omega^2_k$, and the other with the field mode density modified by the resonator~\cite{Heinzen1987}:
\begin{equation}
	\rho_{\text{cav}}(\omega_k) = \begin{cases}
		\rho(\omega_k), &\mathrm{for~} \vb{k} \text{~in~} \Delta \Omega_{\text{cav}},\\
		\rho_{\text{free}}(\omega_k), &\mathrm{for~} \vb{k} \text{~in~} \Delta \Omega_{\text{free}}
	\end{cases},
\end{equation}
where $\rho(\omega_k)$ is given by Eq.~\eqref{dos}, and $\Delta \Omega_{\text{cav}}$ is the solid angle subtended by the resonator mirrors at the center of the cavity. And, the continuum limit on the free space part would have been obtained as $(1/V)\sum_{\vb{k}} \to (2\pi)^{-3} \int \dd{\Omega_{\vb{k}}} \int \dd{\omega_k} \rho_{\mathrm{free}}(\omega_k)$.

The qualitative features of interest to us, that is, the variation of the Lamb shift with cavity detuning, will not change if we consider a concentric resonator in place of the closed cavity. Therefore, for convenience, we work with a closed cavity.

Evaluating the integral in Eq.~\eqref{CalG3}, we obtain
\begin{equation}\label{CalG4}
	\begin{split}
		\mathcal{G}_{22}(\bar{\nu}) &= \frac{\gamma}{3 \pi \hbar \epsilon_0 V}  \Big[\rho(\bar{\nu}) \bar{\nu} \Theta(\bar{\nu}) \\
		&+ \frac{\zeta(\omega)}{4}  \Big[(\bar{\nu}  + \omega) \rho(\bar{\nu}  + \omega) \Theta(\bar{\nu} + \omega)  \\
		&+ (\bar{\nu} - \omega) \rho(\bar{\nu} - \omega) \Theta(\bar{\nu} - \omega)\Big] -  \frac{2}{5} \Big\{ \frac{\bar{\nu}^2 R^2}{c^2} \bar{\nu} \rho(\bar{\nu}) \Theta(\bar{\nu}) \\
		&- \frac{1}{2}  \Big( \frac{(\bar{\nu} + \omega)^2 R^2}{c^2} (\bar{\nu} + \omega) \rho(\bar{\nu} + \omega) \Theta(\bar{\nu} + \omega) \\
		&+ \frac{(\bar{\nu} - \omega)^2 R^2}{c^2} (\bar{\nu} - \omega) \rho(\bar{\nu} - \omega) \Theta(\bar{\nu} - \omega)\Big) \Big\} \Big],
	\end{split}
\end{equation}
where $\Theta(x)$ is the Heaviside theta function.
%%%%%%%%%%%%%%%%%%%%%%%%%%%%%%%%%%%%%%%%%%%%%%%%%%%%%%%%%%%%%%%%%%%%%%%%%%%%%%%%%%%%%%%%%%%%%%%% 
\begin{figure*}
	\centering
	\subfigure[]{
		\includegraphics[width=0.4\linewidth]{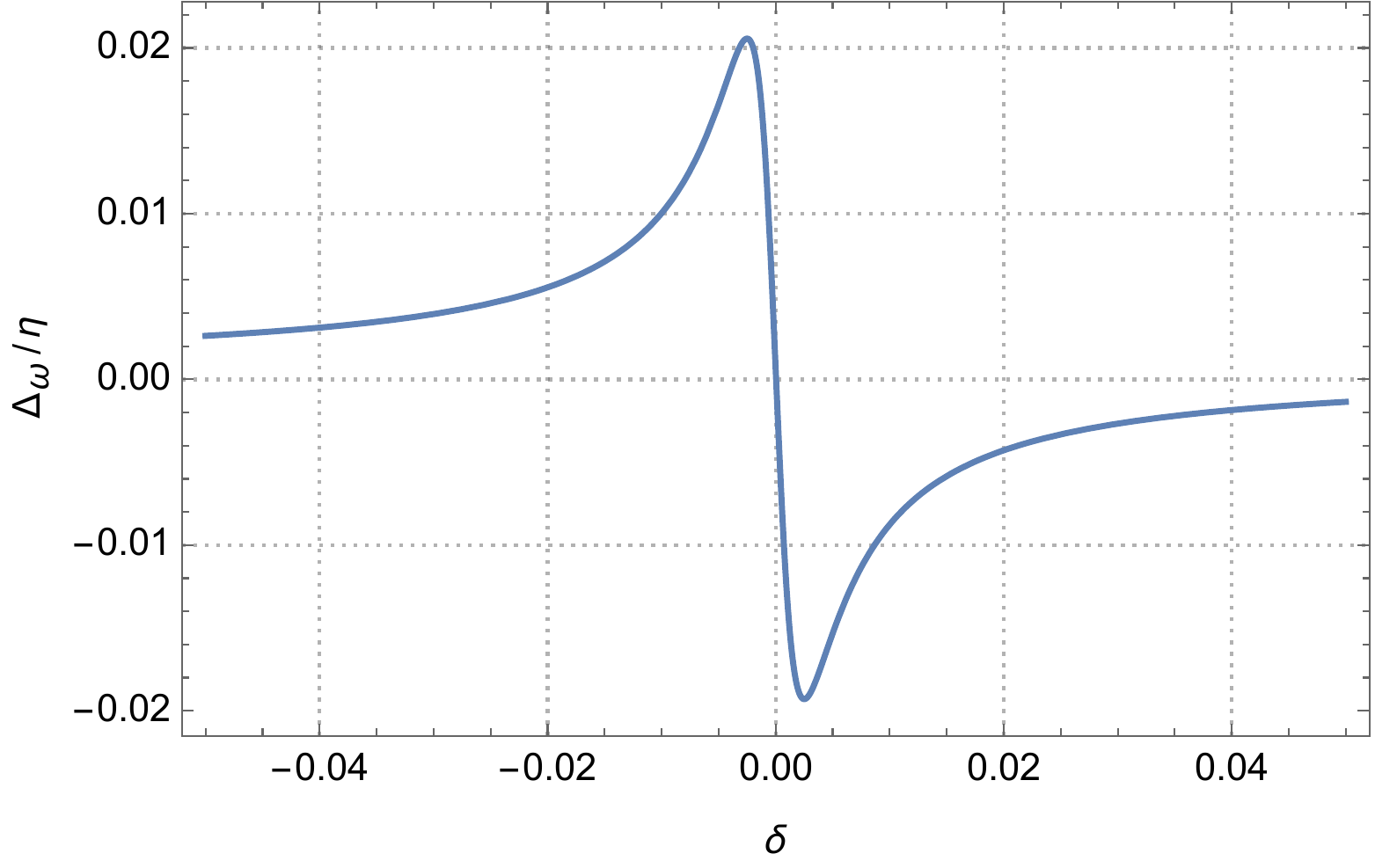}
		\label{fig:LSrot1a}}
	\subfigure[]{
		\includegraphics[width=0.4\linewidth]{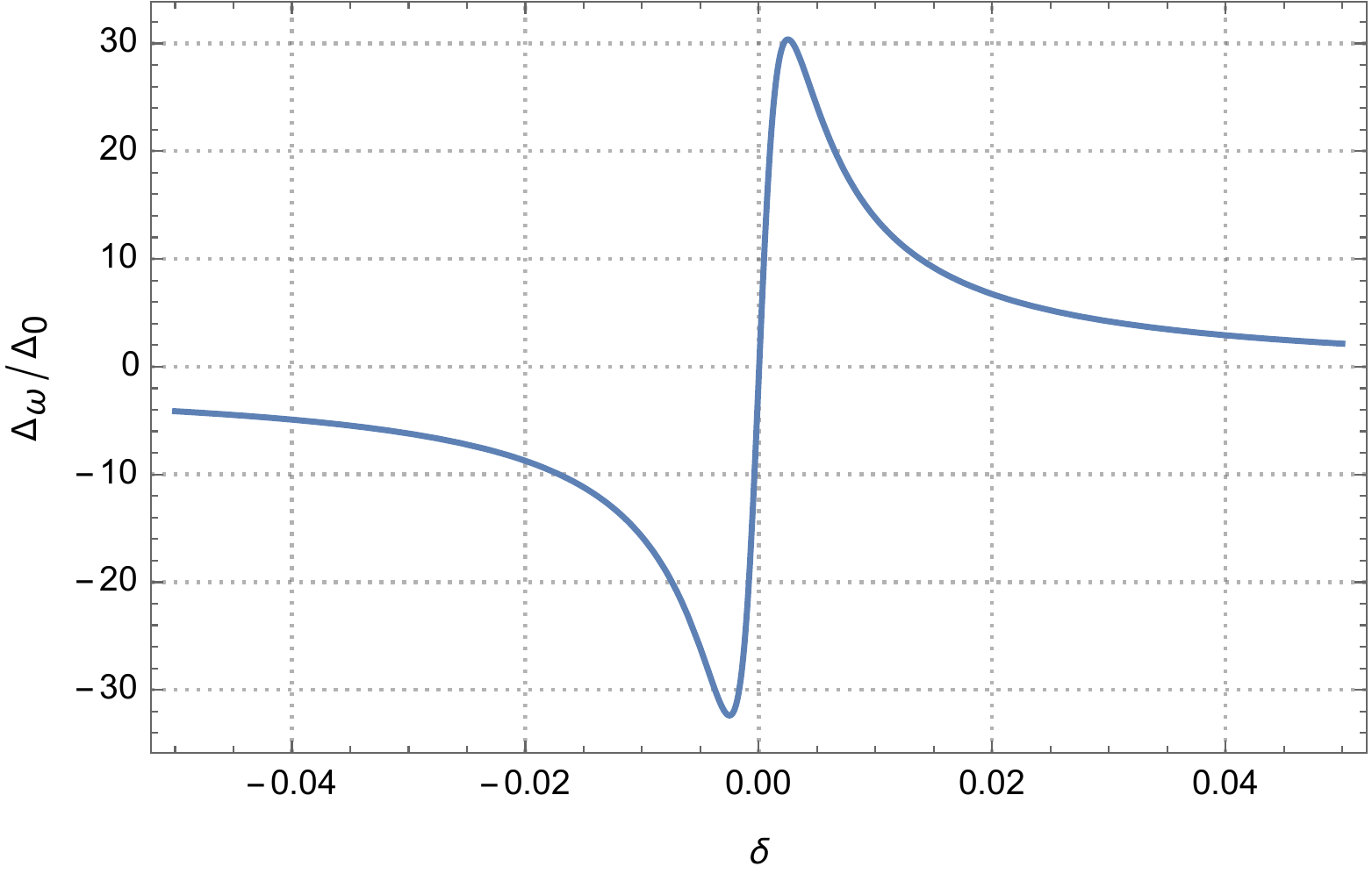}
		\label{fig:LSrot1b}}
	\subfigure[]{
		\includegraphics[width=0.385\linewidth]{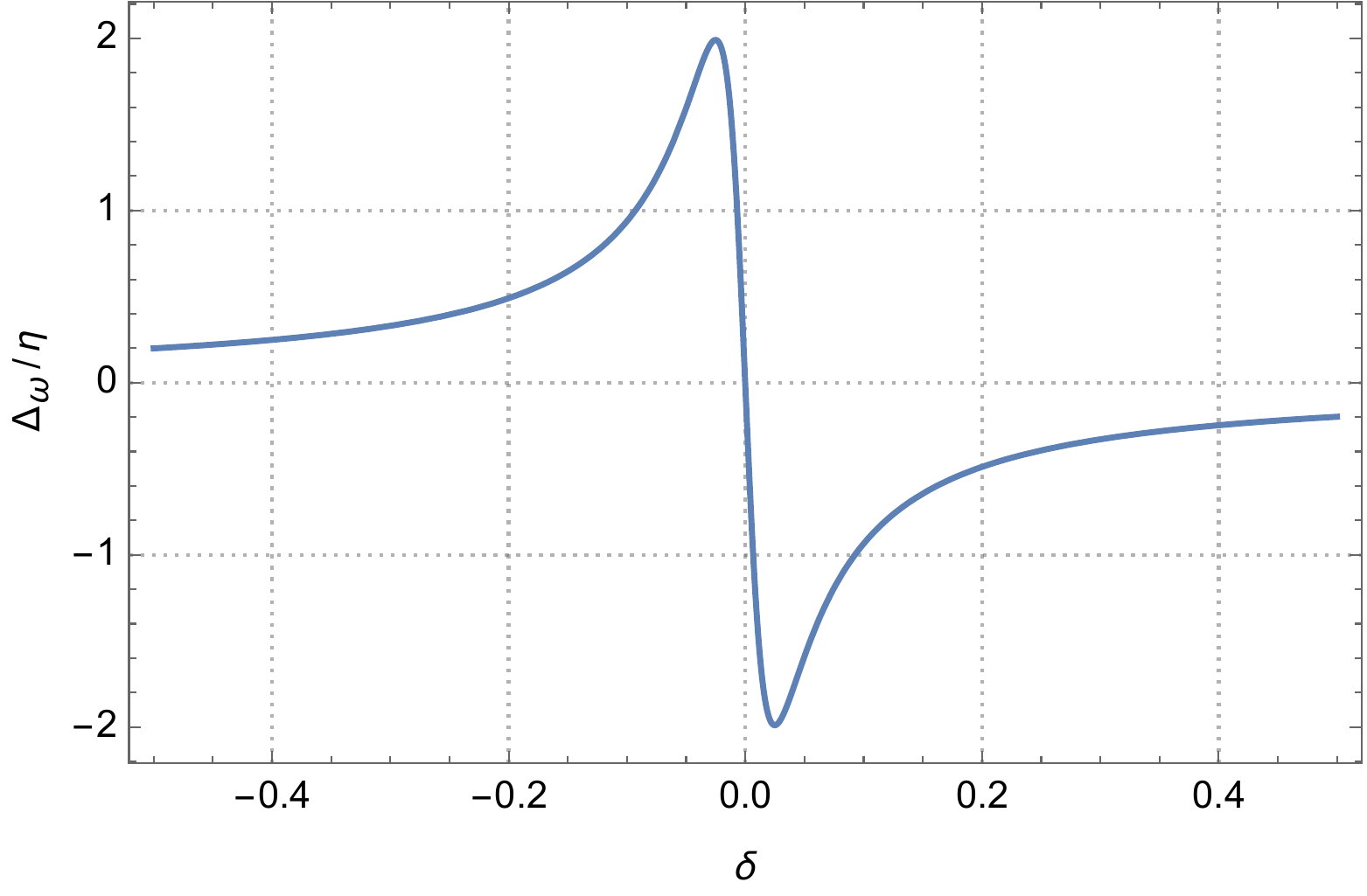}
		\label{fig:LSrot2a}}
	\subfigure[]{
		\includegraphics[width=0.415\linewidth]{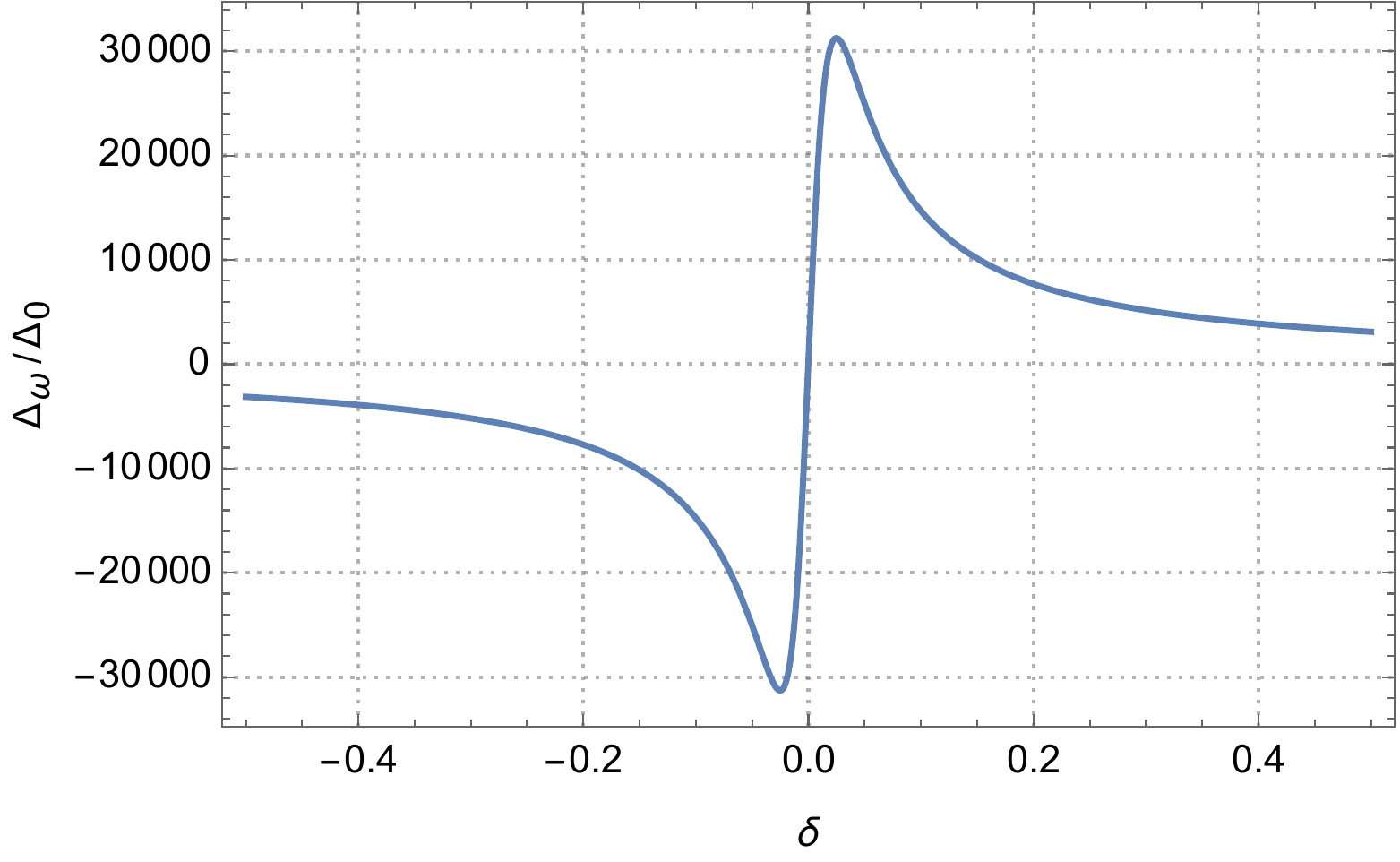}
		\label{fig:LSrot2b}}
	\caption{The $\Delta_{\omega}/\eta$ versus $\omega_{\rm c}$ and $\Delta_{\omega}/\Delta_0$ versus $\omega_{\rm c}$ plots for a two-level system on a circular trajectory of $R = 10^{-5}~\mathrm{m}$ with $\omega = 5 \times 10^{9}~\mathrm{Hz}$ (for plots (a) and (b)) and $\omega = 5 \times 10^{10}~\mathrm{Hz}$ (for plots (c) and (d)). For plots (a) and (c), we have written $\omega_{\rm c} = \omega + \bar{\Omega}_0 + \delta \times 10^5$. For values of cavity's normal frequency in the vicinity of $\omega \pm \bar{\Omega}_0$, large spikes in both $\Delta_{\omega}/\eta$ and  $\Delta_{\omega}/\Delta_0$ are recorded. For clarity, the plots show only the spike at $\omega + \bar{\Omega}_0$ [see Fig.~\eqref{fig:LS(5x10^11)}]. The plots (a) and (b) correspond to an average accelerations of $ a = \omega^2 R \sim 10^{14}~\mathrm{m/s^2}$, while the plots (c) and (d) correspond to an average acceleration of $ a = \omega^2 R \sim 10^{16}~\mathrm{m/s^2}$. For the plots we have taken $Q \sim 10^7$, and $V \sim 10^{-8}~\mathrm{m^3}$ which gives $\eta \sim 10^{-6}~\mathrm{Hz}$. See the discussion following Eq.~\eqref{badcavity} on how to chose cavity parameters consistent with the bad-cavity regime.}
	\label{fig:LS&Ratio}
\end{figure*} 
\subsection{Lamb shift in the circulating atom}
The Lamb shift in the circulating atom can be obtained by combining Eqs.~\eqref{LS1} and \eqref{CalG4}. To obtain the Lamb shift we first recast Eq.~\eqref{LS1} using $\bar{\nu}=\nu/\gamma $ and  $\bar{\Omega}_0=\Omega_0/\gamma $, as
\begin{equation}
	\begin{split}
		\Delta &= \frac{\abs{\vb{d}'}^2}{2 \pi} \int_{-\infty}^{\infty} \dd{\bar{\nu}} \mathcal{G}_{22}( \bar{\nu}) \mathrm{P.V.}\left(\frac{1}{\bar{\nu} + \bar{\Omega}_0} - \frac{1}{\bar{\nu} - \bar{\Omega}_0}\right),
	\end{split}
\end{equation}
which, using Eq.~\eqref{CalG4}, leads to the total Lamb shift given by
\begin{widetext}
\begin{multline}\label{LS2}
	\Delta = \frac{\gamma \eta}{2 \pi} \int_{-\infty}^{\infty} \dd{\bar{\nu}}   \Bigg[\rho(\bar{\nu}) \bar{\nu} \Theta(\bar{\nu}) +\frac{\zeta(\omega)}{4}  \left[(\bar{\nu}  + \omega) \rho(\bar{\nu}  + \omega) \Theta(\bar{\nu} + \omega)  + (\bar{\nu} - \omega) \rho(\bar{\nu} - \omega) \Theta(\bar{\nu} - \omega)\right] \\ 
	-  \frac{2}{5} \left\{ \zeta(\bar{\nu}) \bar{\nu} \rho(\bar{\nu}) \Theta(\bar{\nu}) - \frac{1}{2}  \left( \zeta(\bar{\nu} + \omega) (\bar{\nu} + \omega) \rho(\bar{\nu} + \omega) \Theta(\bar{\nu} + \omega) + \zeta(\bar{\nu} - \omega) (\bar{\nu} - \omega) \rho(\bar{\nu} - \omega) \Theta(\bar{\nu} - \omega)\right) \right\} \Bigg] \\
	\times \mathrm{P.V.}\left(\frac{1}{\bar{\nu} + \bar{\Omega}_0} - \frac{1}{\bar{\nu} - \bar{\Omega}_0}\right),
\end{multline}
\end{widetext}
\begin{figure}
	\centering
	\subfigure[]{
		\includegraphics[width=0.415\linewidth]{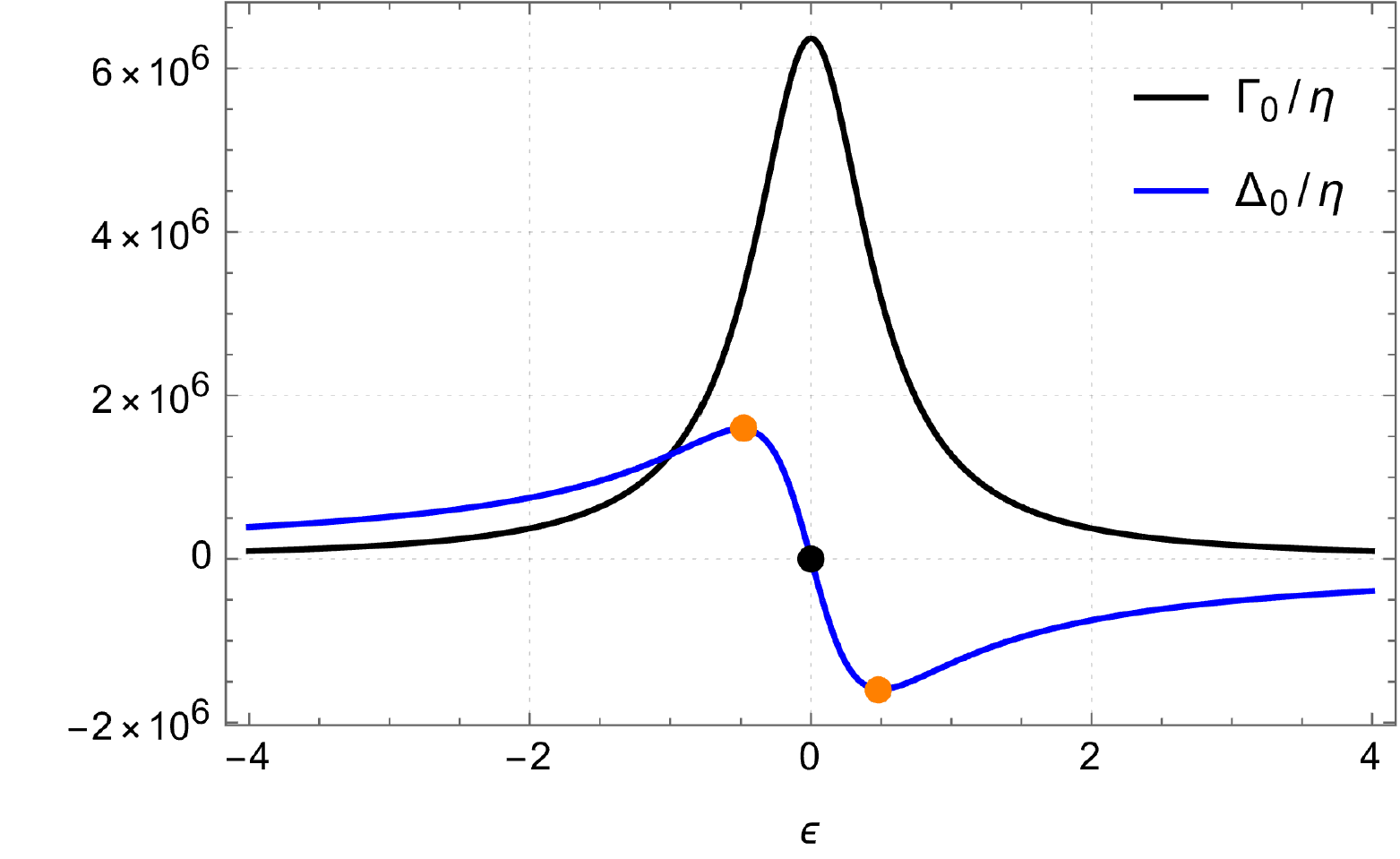}
		\label{fig:LW&LSin}}
	\subfigure[]{
		\includegraphics[width=0.385\linewidth]{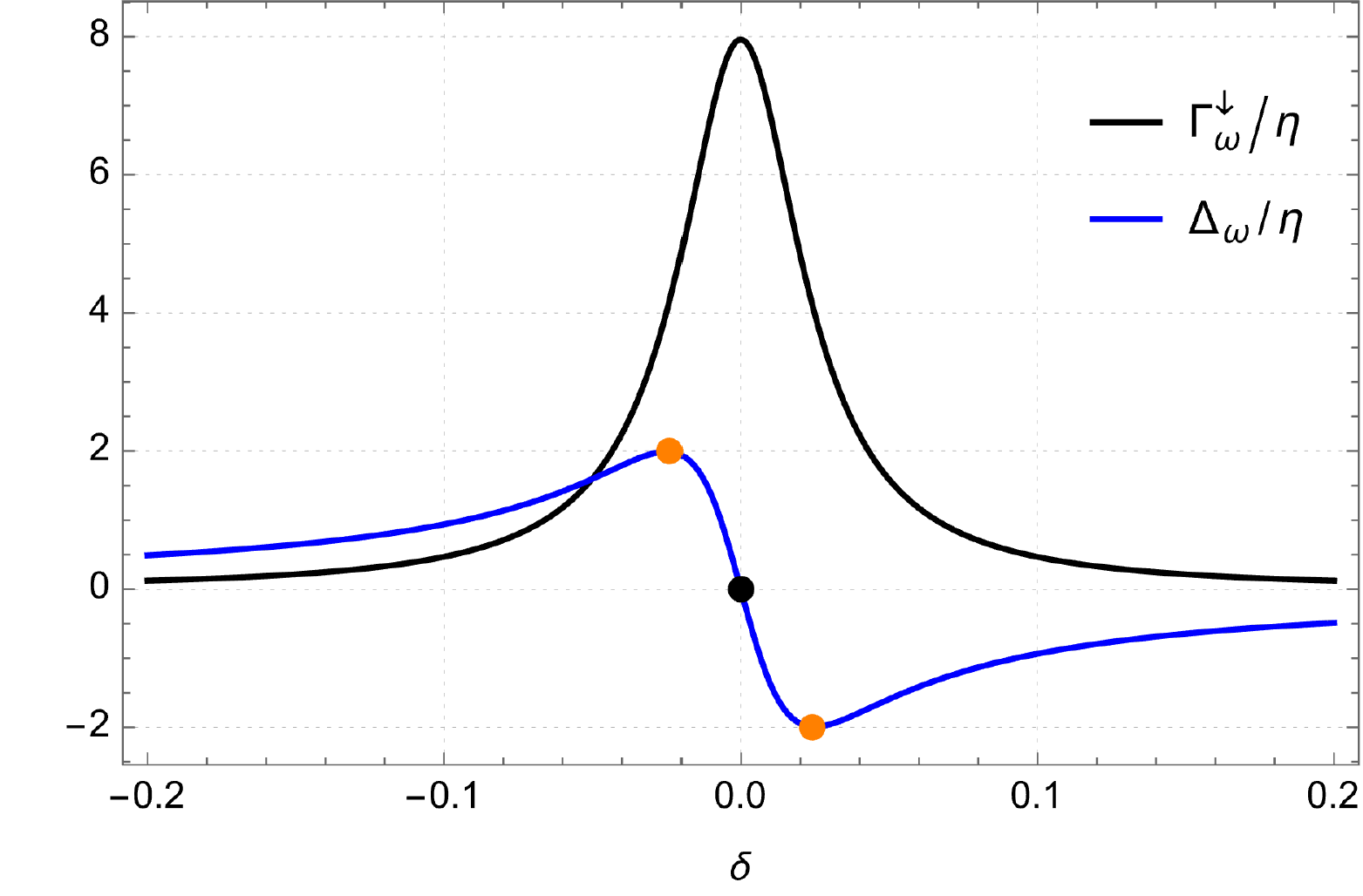}
		\label{fig:LW&LSrot}}
	\caption{Plot~(a) shows the decay rate $\Gamma_0 = \eta \rho(\Omega_0) \Omega_0$~\cite{lochan2020,AryaPRD2022}, and Lamb shift $\Delta_0$~[Eq.~\eqref{LSrest}] of an inertial atom with $\Omega_0 = 10$MHz, as a function of the cavity's normal frequency $\omega_{\rm c} = \Omega_0 + \epsilon$. The general behavior of the two quantities w.r.t cavity detuning agrees well with experimental results~\cite{Heinzen1987}. Similarly, plot~(b) shows the purely-noninertial decay rate $\Gamma_{\omega}^{\downarrow}$~[Eq.~\eqref{Gammarot1}] and Lamb shift $\Delta_{\omega}$~[Eq.~\eqref{NonInLS}] of a circulating atom (with $\Omega_0 = 10$MHz and $\omega = 50$GHz) as a function of the cavity's normal frequency $\omega_{\rm c} = \omega + \Omega_0 + \delta \times 10^5$. In the plots, a black dot marks the point at which the Lamb shift vanishes, and orange dots mark the points at which the Lamb shift attains its minimum or maximum value.}
	\label{fig:LW&LS}
\end{figure}
where $\eta \equiv \abs{\vb{d}' }^2/(3 \pi \hbar \epsilon_0 V)$, $\zeta(\nu) \equiv \nu^2 R^2/c^2$. Eq.~(28) shows the dependence of the Lamb shift on the atom's acceleration $a = \omega^2 R$ through the parameter $\zeta(\omega) \equiv \omega^2 R^2/c^2$.
The parameter $\eta$ has the dimensions of inverse time and can be expressed in terms of the atom-cavity coupling constant $g \equiv \abs{\vb{d}'} \sqrt{\omega_{c}/(2 \hbar \epsilon_0 V)}$, and cavity's normal frequency $\omega_{\rm c}$ as $\eta = 2g^2/3\pi \omega_{\rm c}$. The volume dependence of $\Delta$ is entirely contained in $\eta$. To obtain plots independent of a specific choice of the mode volume $V$ of the cavity, we are plotting $\Delta_0/\eta$ and $\Delta_{\omega}/\eta$ as a function of the cavity's normal frequency $\omega_{\rm c}$. For different values of the atom's angular frequency, we take different values of $V$ for an optimum signal while maintaining consistency with the bad-cavity regime mentioned in Eq.~(13). The captions of Figs.~(1) and (2) mention the corresponding values of $V$ and $\eta$. These $\eta$ values are then used to determine the purely-noninertial and inertial contributions to the Lamb shift.

From Eq.~\eqref{LS2}, the Lamb shift of an inertial atom can be obtained in the limit $\omega \rightarrow 0$.  For the inertial Lamb shift we obtain
\begin{equation}\label{LSrest}
		\Delta_{0} = \frac{\eta}{2 \pi} \int_{-\infty}^{\infty} \dd{\nu} \rho(\nu) \nu \Theta(\nu) \mathrm{P.V.}\left(\frac{1}{\nu + \Omega_0} - \frac{1}{\nu - \Omega}_0\right).
\end{equation}
The Cauchy Principal value integral can be evaluated exactly. For a high-$Q$ cavity, specifically, we obtain a simpler expression given by~[see Appendix~\ref{ap:Inertial Lamb shift}]
\begin{equation}
	\Delta_{0} \approx  - \eta \Omega_0 \left[\frac{\omega_c (\Omega_0^2 + \omega_c^2) \log(\omega_c/\Omega_0)}{2 \pi^2 Q  \left(  \omega_c^2 -  \Omega_0^2\right)^2} + \frac{\omega_c}{\pi  \left(  \omega_c^2 -  \Omega_0^2\right)}  \right],
\end{equation}
for $\omega_c \neq \Omega_0$. As can be seen in Fig.~\eqref{fig:LS(rest)}, $\Delta_{0}$ maximizes when the cavity is tuned near, but not exactly at, the atomic resonance $\Omega_0$.
The purely-noninertial contribution to the Lamb shift can be obtained as 
\begin{equation}\label{NonInLS}
	\Delta_{\omega} = \Delta - \Delta_{0}.
\end{equation}
The three integrals in the first line of Eq.~\eqref{LS2} are manifestly convergent. The three integrals in the second line, however, are individually logarithmically divergent but the total Lamb shift is convergent as shown in Appendix~\ref{ap:Convergence}. One can obtain a closed form expression for $\Delta_{\omega}$, however, the expression is cumbersome without offering additional insights. Therefore, we instead resort to the plots [Figs.~\eqref{fig:LS(5x10^11)} and \eqref{fig:LS&Ratio}] to illustrate the main features.

Note the presence of $\zeta(\bar{\nu} \pm \omega) (\bar{\nu} \pm \omega) \rho(\bar{\nu} \pm \omega) \Theta(\bar{\nu} \pm \omega)$ terms in Eq.~\eqref{LS2}. The plots show that the purely-noninertial contribution to the Lamb shift, in contrast with the inertial contribution, maximizes away from the atomic resonance $\Omega_0$, in the vicinity of frequencies $\omega \pm \bar{\Omega}_0$. This allows us to isolate and enhance the purely-noninertial contribution relative to the inertial contribution by appropriately tuning the cavity. 

At this point, recall that the radiative shift to the energy levels of an atom occurs through processes mediated by virtual photons. For these processes, the state of the radiation field remains unchanged in the initial and final states of the atom-field composite system, and changes only in the intermediate composite state~\cite{API_Tannoudji}. These processes complement the processes in which the emission or absorption of a real photon is involved which lead to transitions between different energy levels of the atom. Fig.~3(a) shows the emission rate and Lamb shift in an inertial atom as a function of the cavity's normal frequency $\omega_{\rm c}$. An inertial two-level atom undergoes transitions dominantly at $\Omega_0$, while the Lamb shift in its energy levels maximizes when the cavity is tuned to a frequency in the vicinity of $\Omega_0$ as shown in Fig.~\ref{fig:LS(rest)}.
As was reported and exploited in refs.~\cite{lochan2020,AryaPRD2022}, a circulating two-level atom undergoes transitions at field frequencies $\omega_k = \abs{\omega \pm \bar{\Omega}_0}$, in addition to the transitions at the inertial resonant frequency $\Omega_0$. The purely-noninertial decay rate of the circulating atom is given to the first order in $\zeta(\omega)$ by~\cite{lochan2020,AryaPRD2022} 
		\begin{equation}\label{Gammarot1}
			\begin{split}
				\Gamma_{\omega}^{\downarrow} &= \frac{\eta \zeta(\omega)\Omega_0}{2}  \Bigg[- \Omega_0 \rho'(\Omega_0) + \frac{9}{10}  \frac{\omega + \bar{\Omega}_0}{\Omega_0} \rho(\omega + \bar{\Omega}_0) \Bigg]
			\end{split}
		\end{equation}
		where $\rho'(\omega_k) = \partial \rho / \partial \omega_k$. Fig.~3(b) depicts $\Gamma^{\downarrow}_{\omega}$ and $\Delta_{\omega}$ as a function of the cavity's normal frequency in the neighborhood of $\omega + \bar{\Omega}_0$.
		
Thus we see that for a microwave two-level system, weakly coupled to EM field inside a bad-cavity, when the cavity is tuned at the atomic resonance, a maximum inertial Lamb shift of the order of $10^{-3}~\mathrm{Hz}$ is obtained. Further, by tuning the cavity in the vicinity of either $\omega + \bar{\Omega}_0$ or $\omega - \bar{\Omega}_0$, the purely-noninertial contribution can be made dominant~[see Figs.~\eqref{fig:LS(5x10^11)} and \eqref{fig:LS&Ratio}]. For example, at an average acceleration $a = \omega^2 R \sim 10^{14}~\mathrm{m/s^2}$, a noninertial contribution $\Delta_{\omega}\sim 10^{-8}~\mathrm{Hz}$, and for $a \sim 10^{16}~\mathrm{m/s^2}$, a purely-noninertial contribution $\Delta_{\omega} \sim 10^{-6}~\mathrm{Hz}$ to the Lamb shift can be obtained. Further, at accelerations achievable with electrons inside storage rings~\cite{bell1983}, a purely-noninertial contribution as large as $\sim 10^{-3}~\mathrm{Hz}$ can be obtained. The plots \ref{fig:LSrot1b} and \ref{fig:LSrot2b} show the enhancement $\Delta_{\omega}/\Delta_0$ of the purely-noninertial contribution to the Lamb shift over the inertial contribution. This enhancement signifies the degree of decontamination of the Lamb shift from the inertial contribution in the sense that the interest of some prospective experiment would be in the purely-noninertial contribution.
Given that the Lamb shift in Hydrogen atom has been measured to seven significant digits~\cite{Bezginov2019}, the purely-noninertial contribution to the Lamb shift can be observed with current~\cite{Ahn2018} or near-future technology.

\section{Discussion and Conclusions}\label{sec:discussion}
In this work, we have studied the radiative energy shifts in a first-quantized two-level system on a uniform circular trajectory due to its interaction with a quantum electromagnetic field inside a cavity. We have argued that the radiative energy shift in atomic levels is an observable of interest for the detection of noninertial effects. We have shown that the inertial contribution to the Lamb shift in a circulating atom maximizes in the vicinity of the atomic resonance whereas the purely-noninertial contribution maximizes away from the atomic resonance at frequencies decided by the atom's angular frequency. By suitably tuning the cavity parameters, an observable purely-noninertial contribution $\sim 10^{-8}-10^{-6}~\mathrm{Hz}$ can be obtained for average accelerations $\sim 10^{14} - 10^{16}~\mathrm{m/s^2}$.

It is instructive to compare the current theoretical proposal with other proposals aiming at the detection of the noninertial effects. In free space, the Unruh effect demands acceleration of the order of $10^{21} \mathrm{m/s^2}$ if the detector transition rate~\cite{unruh1976} or radiative energy shift~\cite{Audretsch1995PRA} is observed, and $10^{17}\mathrm{m/s^2}$ if the observed quantity is geometric phase~\cite{Martinez2011}. Inside a long cylindrical cavity, however, it has been argued that the Unruh effect can be detected at accelerations as low as $\sim 10^{9} \mathrm{m/s^2}$ by observing the atom's spontaneous emission. Similarly, detection of noninertial effects due to uniform circular motion of an atom inside an electromagnetic cavity, by observing the atomic spontaneous decay rate, requires an acceleration of the order of $10^{14} \mathrm{m/s^2}$~\cite{lochan2020}. Whereas, if the geometric phase is observed, such noninertial effects may become detectable at accelerations as low as $\sim 10^{7} \mathrm{m/s^2}$~\cite{AryaPRD2022}.
This comparison illustrates the role of the density of field states in relation to the amplification of the noninertial effects. In addition, this comparison points to the fact that different system properties might differ in their ability to capture such effects, which can possibly guide the selection and design of suitable experiments. For example, since the spectroscopic techniques are much well-established as compared to the mixed-state geometric phase measurements, the Lamb shift proposal might be easier to implement than the geometric phase proposal.

Furthermore, the current work can be generalized in several directions. For example, we have considered only a two-level atom but real atoms have more than two levels which can lead to new features~\cite{Wilson2003}. Also note that we have worked in the weak-coupling regime. An analysis involving strong coupling between the atom and the field can possibly lead to even higher energy shifts. 

Additionally, the results obtained here encourage one to investigate the Lamb shift for the detection of the Unruh effect~\cite{unruh1976}. A cavity with a cylindrical geometry suits the requirement of uniform linear acceleration in the case of the Unruh effect and has been argued to possibly facilitate its detection at low accelerations when the atomic transition rates are observed~\cite{Jaffino2022}. The problem of the possibility of detecting the Unruh effect through the radiative energy shifts in an atom undergoing uniform linear acceleration inside a cylindrical cavity will be taken up in a subsequent work.
\begin{acknowledgments}
N.A. acknowledges financial support from the University Grants Commission (UGC), Government of India, in the form of a research fellowship (Sr.~No.~2061651285). N.A. thanks Kinjalk Lochan and Vikash Mittal for carefully reading the manuscript and making useful suggestions.
\end{acknowledgments}

\appendix
\section{Derivation of Eq.~\eqref{CalG2a}}\label{ap:crossterms}
In Eq.~\eqref{CalG2b}, introducing a change of integration variables $(t_1,t_2) \mapsto (t_+, t_-)$ with
\begin{subequations}\label{t+t-}
	\begin{align}
		t_+ &\equiv (t_1 + t_2)/2,\\
		t_- &\equiv t_2 - t_1,
	\end{align}
\end{subequations}
and using Eq.~\eqref{E_2}, we obtain
\begin{equation}\label{CalG1}
	\begin{split}
		&\mathcal{G}'_{22}(\nu) = \lim_{T \rightarrow \infty} \frac{1}{\hbar^2 T} \int_{0}^{\gamma T} \dd{t_+} \int_{-\gamma T}^{\gamma T} \dd{t_-} e^{i \bar{\nu} t_-} \times \\
		&\bra{0} \left\{E^{y}(x^{\mu}_2)- R\omega \left( \sin(\omega t_2)B^{z}(x^{\mu}_2) - \cos(\omega t_2) B^{x}(x^{\mu}_2) \right)\right\} \\
		&\times \left\{E^{y}(x^{\mu}_1)- R\omega \left(\sin(\omega t_1)B^{z}(x^{\mu}_1) - \cos(\omega t_1) B^{x}(x^{\mu}_1) \right) \right\} \ket{0}.
	\end{split}
\end{equation}
Next, we show that in Eq.~\eqref{CalG1} the contribution of the terms containing cross-correlations of the form $\bra{0}E^i(x^{\mu}_2)B^j(x^{\mu}_1)\ket{0}$ and $\bra{0}B^k(x^{\mu}_2)B^l(x^{\mu}_1)\ket{0}, k \neq l$; vanishes. 
Collecting the cross-terms in \eqref{CalG1}, and denoting the collection by $\mathcal{G}_{22}^{'\text{CT}}$, we get
\begin{multline}\label{G_CT}
	\begin{split}
		\mathcal{G}^{'\text{CT}}_{22}(\nu) &= \lim_{T \rightarrow \infty} \frac{1}{T} \int_{0}^{\gamma T} \dd{t_+} \int_{-\gamma T}^{\gamma T} \dd{t_-} e^{i \bar{\nu} t_-} \\
		&\Big[ -R\omega \Big\{\sin(\omega t_2) \bra{0} B^{z}(x^{\mu}_2)E^{y}(x^{\mu}_1)\ket{0}\\
		&+ \sin(\omega t_1) \bra{0} E^{y}(x^{\mu}_2) B^{z}(x^{\mu}_1)\ket{0} \\
		&+ \cos(\omega t_2) \bra{0}B^{x}(x^{\mu}_2)E^{y}(x^{\mu}_1)\ket{0}   \\
		&+ \cos(\omega t_1) \bra{0} E^{y}(x^{\mu}_2) B^{x}(x^{\mu}_1)\ket{0} \Big\} \\
		&+ R^2\omega^2 \Big\{\cos(\omega t_2) \sin(\omega t_1)  \bra{0}B^{x}(x^{\mu}_2) B^{z}(x^{\mu}_1)\ket{0} \\
		&+ \sin(\omega t_2) \cos(\omega t_1)\bra{0} B^{z}(x^{\mu}_2) B^{x}(x^{\mu}_1)\ket{0} \Big\} \Big]\\
		&\equiv I_1 + I_2 + I_3 + I_4 + I_5 + I_6.
	\end{split}
\end{multline}
It is straightforward to conclude that $I_1 + I_2$ and $I_3 + I_4$ vanish under $t_+-$integration. For the remaining two terms, we have
\begin{equation}
	\begin{split}
		I_5 + I_6 &\equiv R^2\omega^2 \lim_{T \rightarrow \infty} \frac{1}{T} \int_{0}^{\gamma T} \dd{t_+} \int_{-\gamma T}^{\gamma T} \dd{t_-} e^{i\bar{\nu}t_-} \\
		& \times \Bigg[ \cos(\omega t_2) \sin(\omega t_1) \bra{0}B^x(x_2) B^z(x_1) \ket{0} \\
		&+ \sin(\omega t_2) \cos(\omega t_1) \bra{0}B^z(x_2) B^x(x_1) \ket{0}\Bigg]\\
		&= \frac{R^2\omega^2}{2} \lim_{T \rightarrow \infty} \frac{1}{T} \int_{0}^{\gamma T} \dd{t_+} \Bigg[-\int_{-\gamma T}^{\gamma T} \dd{t_-} e^{i\bar{\nu} t_-} \\
		& \times \sin(\omega t_-) \bra{0} B^x(x_2) B^z(x_1) \ket{0}  + \int_{-\gamma T}^{\gamma T} \dd{t_-} e^{i\bar{\nu} t_-} \\
		& \times \sin(\omega t_-) \bra{0} B^z(x_2) B^x(x_1) \ket{0}\Bigg],
	\end{split}
\end{equation}
where we have used Eqs.~\eqref{t+t-} and the terms containing $\sin(2\omega t_+)$ vanished under the $t_+-$integration.
Further, Eq.~\eqref{B} and
\begin{equation}
	\sum_{\lambda}\left(\vu{k} \times \epsilon_{\vb{k}\lambda}\right)^i  \left(\vu{k} \times \epsilon_{\vb{k}\lambda}\right)^j = \delta^{ij}-\frac{k^i k^j}{k^2},
\end{equation}
lead to
\begin{equation}\label{BiBj}
	\begin{split}
	\bra{0} {B}^i(x^{\mu}_2) {B}^j(x^{\mu}_1) \ket{0} &=  \frac{1}{c^2} \sum_{\vb{k}} \frac{\hbar \omega_k}{2\epsilon_0 V} \left(1-\frac{k^i k^j}{k^2}\right) \\
	& \times e^{-i\omega_k (t_2 - t_1)} e^{i \vb{k} \cdot (\vb{x}(t_2) - \vb{x}(t_1))}.
	\end{split}
\end{equation}
From Eq.~\eqref{BiBj} we have $\bra{0} {B}^i(x^{\mu}_2) {B}^j(x^{\mu}_1) \ket{0} = \bra{0} {B}^j(x^{\mu}_2) {B}^i(x^{\mu}_1) \ket{0}$ which leads to $I_5 + I_6 = 0$. Thus, we obtain
\begin{multline}\label{CalG2}
	\begin{split}
		\mathcal{G}'_{22}(\nu) &= \lim_{T \rightarrow \infty} \frac{1}{\hbar^2 T} \int_0^{\gamma T} \dd{t_+} \int_{-\gamma T}^{\gamma T} \dd{t_-} e^{i \bar{\nu} t_-} \\ 
		& \times \Big[\bra{0}E^{y}(x^{\mu}_2) E^{y}(x^{\mu}_1) \ket{0} \\
		&+ \frac{R^2\omega^2}{2} \cos(\omega t_-)\bra{0} B^{z}(x^{\mu}_2) B^{z}(x^{\mu}_1) \ket{0} \\
		&+ \frac{R^2\omega^2}{2} \cos(\omega t_-) \bra{0}B^{x}(x^{\mu}_2) B^{x}(x^{\mu}_1) \ket{0}  \Big].
	\end{split}
\end{multline}
Finally, performing the $t_+$-integral leads to Eq.~\eqref{CalG2a}.
%%%%%%%%%%%%%%%%%%%%%%%%%%%%%%%%%%%%%%%%%%%%%%%%%%%%%%%%%%%%%%%%%%%%%%%%%%%%%%%%%%%%%%%%%%%%%%%%%%%%%%%%%%%%%%%
\section{Inertial Lamb shift}\label{ap:Inertial Lamb shift}
Performing the integral in Eq.~\eqref{LSrest}, we obtain
\begin{equation}\label{LSrest1}
	\begin{split}
		&\Delta_{0} = \frac{\eta}{2\pi^2  \left(8 Q^2 \left(1-4 Q^2\right) \Omega _0^2 \omega _c^2+\left(4 Q^2+1\right)^2 \omega _c^4+16 Q^4 \Omega _0^4\right)} \\
		& \times \Bigg[8 Q^3 \Omega _0^3 \omega _c \Bigg\{\log \left(\frac{4 Q^2}{\left(4 Q^2+1\right) \omega _c^2}\right)+ 2 Q \left( \pi + 2 \tan ^{-1}(2 Q)\right) \\
		&+ 2 \log \left(\Omega _0\right)\Bigg\} - 2 Q \left(4 Q^2+1\right) \Omega _0 \omega _c^3 \Bigg\{2 \log \left(\omega _c\right) \\
		&+\log \left(\frac{1}{4 Q^2}+1\right)+ 2 Q \left( \pi + 2 \tan ^{-1}(2 Q)\right) - 2 \log \left(\Omega _0\right)\Bigg\}\Bigg].
	\end{split}
\end{equation}
For $Q \gg 1$, we can ignore $1$ in comparison to $Q$. Also, using $\lim_{Q \to \infty} \tan^{-1}Q = \pi/2$, we can write
\begin{multline}
		\Delta_{0} \approx \frac{\Omega_0 \omega_c \eta }{4 \pi^2 Q  \left(  \omega_c^2 -  \Omega_0^2\right)^2} \Bigg[- (\Omega_0^2 + \omega_c^2) \log(\omega^2_c) \\
		+ 4 \pi Q (\omega_0^2 - \omega_c^2) + (\Omega_0^2 + \omega_c^2) \log(\Omega^2_0) \Bigg].
\end{multline}
Further rearrangements and simplifications lead to
\begin{equation}
	\Delta_{0} \approx  - \eta \Omega_0 \left[\frac{\omega_c (\Omega_0^2 + \omega_c^2) \log(\omega_c/\Omega_0)}{2 \pi^2 Q  \left(  \omega_c^2 -  \Omega_0^2\right)^2} + \frac{\omega_c}{\pi  \left(  \omega_c^2 -  \Omega_0^2\right)}  \right],
\end{equation}
for $\omega_c \neq \Omega_0$.
%%%%%%%%%%%%%%%%%%%%%%%%%%%%%%%%%%%%%%%%%%%%%%%%%%%%%%%%%%%%%%%%%%%%%%%%%%%%%%%%%%%%%%%%%%%%%%%%%%%%%%%%%%%%%%%%
\section{Convergence of the Lamb shift}\label{ap:Convergence}
Here, we show that the Lamb shift given in Eq.~\eqref{LS2} converges without introducing any ultraviolet cutoff on $\bar{\nu}$. The first three integrals in Eq.~\eqref{LS2} are manifestly convergent. Consider the last three integrals:
\begin{equation}
	\begin{split}
		\mathcal{I} &\equiv \frac{1}{2} \Bigg[\int_{0}^{\infty} \dd{\bar{\nu}} \left\{\frac{\bar{\nu}^2 R^2}{c^2} \bar{\nu} \rho(\bar{\nu}) - \frac{(\bar{\nu} + \omega)^2 R^2}{c^2} (\bar{\nu} + \omega) \rho(\bar{\nu} + \omega) \right\} \\
		& + \int_{0}^{\infty} \dd{\bar{\nu}} \left\{\frac{\bar{\nu}^2 R^2}{c^2} \bar{\nu} \rho(\bar{\nu}) - \frac{(\bar{\nu} - \omega)^2 R^2}{c^2} (\bar{\nu} - \omega) \rho(\bar{\nu} - \omega) \right\}\\
		&- \int_{-\omega}^{0} \dd{\bar{\nu}} \frac{(\bar{\nu} + \omega)^2 R^2}{c^2} (\bar{\nu} + \omega) \rho(\bar{\nu} + \omega)  \\
		&- \int_{0}^{\omega} \dd{\bar{\nu}} \frac{(\bar{\nu} - \omega)^2 R^2}{c^2} (\bar{\nu} - \omega) \rho(\bar{\nu} - \omega) \Bigg]  \\
		& \times \mathrm{P.V.}\left(\frac{1}{\bar{\nu} + \bar{\Omega}_0} - \frac{1}{\bar{\nu} - \bar{\Omega}_0}\right)\\
		&\equiv \mathcal{I}_1 + \mathcal{I}_2 + \mathcal{I}_3 + \mathcal{I}_4,
	\end{split}
\end{equation}
where we have grouped different integrals in a way that is helpful in ascertaining their convergence.
Noting that $\mathcal{I}_3 $ and $\mathcal{I}_4$ are manifestly convergent, we focus on $\mathcal{I}_1$ and $\mathcal{I}_2$:
\begin{equation}
	\begin{split}
		\mathcal{I}_1 &= \frac{1}{2}\int_{0}^{\infty} \dd{\bar{\nu}} \left\{\frac{\bar{\nu}^3 R^2}{c^2} \rho(\bar{\nu}) - \frac{(\bar{\nu} + \omega)^3 R^2}{c^2} \rho(\bar{\nu} + \omega) \right\} \\
		& \times \mathrm{P.V.}\left(\frac{1}{\bar{\nu} + \bar{\Omega}_0} - \frac{1}{\bar{\nu} - \bar{\Omega}_0}\right) \\
		&\equiv \mathcal{I}_{1a} - \mathcal{I}_{1b},
	\end{split}
\end{equation}
where we have defined
\begin{equation}
	\begin{split}
		\mathcal{I}_{1a} &\equiv \frac{\mathrm{P.V.}}{2} \int_{0}^{\infty} \dd{\bar{\nu}} \frac{\bar{\nu}^3 R^2}{c^2} \rho(\bar{\nu})  \left(\frac{1}{\bar{\nu} + \bar{\Omega}_0} - \frac{1}{\bar{\nu} - \bar{\Omega}_0}\right)\\
		&\equiv \frac{R^2}{2c^2} \mathrm{P.V.} \int_{0}^{\infty} \dd{\bar{\nu}} f_{1a}(\bar{\nu},\bar{\Omega}_0),
	\end{split}
\end{equation}
and
\begin{equation}
	\begin{split}
	\mathcal{I}_{1b} &\equiv \frac{\mathrm{P.V.}}{2}\int_{0}^{\infty} \dd{\bar{\nu}} \frac{(\bar{\nu} + \omega)^3 R^2}{c^2} \rho(\bar{\nu} + \omega) \left(\frac{1}{\bar{\nu} + \bar{\Omega}_0} - \frac{1}{\bar{\nu} - \bar{\Omega}_0}\right)\\
	&\equiv \frac{R^2}{2c^2} \mathrm{P.V.} \int_{0}^{\infty} \dd{\bar{\nu}} f_{1b}(\bar{\nu},\bar{\Omega}_0).
	\end{split}
\end{equation}
Let us first analyze $\mathcal{I}_{1a}$: 
\begin{equation}
	\begin{split}
		&\mathcal{I}_{1a} = \frac{R^2}{2 c^2} \mathrm{P.V.} \int_{0}^{\infty} \dd{\bar{\nu}} f_{1a}(\bar{\nu},\bar{\Omega}_0) \\
		&\equiv \frac{R^2}{2 c^2} \left[\mathrm{P.V.} \int_{0}^{\bar{\Omega}_0 + \epsilon} \dd{\bar{\nu}} f_{1a}(\bar{\nu},\bar{\Omega}_0) + \int_{\bar{\Omega}_0 + \epsilon}^{\infty} \dd{\bar{\nu}} f_{1a}(\bar{\nu},\bar{\Omega}_0)\right],
	\end{split}
\end{equation}
where $\epsilon$ is some positive number.
Note that for $\bar{\nu} \rightarrow \infty$, $f_{1a}(\bar{\nu},\bar{\Omega}_0)$ behaves as
\begin{equation}
	f_{1a}(\bar{\nu},\bar{\Omega}_0) \rightarrow - \frac{\bar{\Omega}_0 \omega_c}{\pi Q \bar{\nu}} - \frac{2 \bar{\Omega}_0  \omega^2_c}{\pi Q \bar{\nu}^2} + \order{\frac{1}{\bar{\nu}^3}},
\end{equation}
which means that the integral $\int_{\bar{\Omega}_0 + \epsilon}^{\infty} \dd{\bar{\nu}} f_{1a}(\bar{\nu},\bar{\Omega}_0)$ diverges logarithmically. Therefore, $\mathcal{I}_{1a}$ diverges logarithmically.
Similarly, since for $\bar{\nu} \rightarrow \infty$ the integrand $f_{1b}(\bar{\nu},\bar{\Omega}_0)$ of $\mathcal{I}_{1b}$ behaves as
\begin{equation}
	f_{1b}(\bar{\nu},\bar{\Omega}_0) \rightarrow  - \frac{\bar{\Omega}_0 \omega_c}{\pi Q \bar{\nu}} - \frac{\bar{\Omega}_0 \omega_c (\omega + 2 \omega_c)}{\pi Q \bar{\nu}^2} + \order{\frac{1}{\bar{\nu}^3}},
\end{equation}
$\mathcal{I}_{1b}$ diverges logarithmically as well. But $\mathcal{I}_{1}$, being the difference of $\mathcal{I}_{1a}$ and $\mathcal{I}_{1b}$, tends to zero faster than $1/\bar{\nu}$ and hence converges. Similarly we can deduce that $\mathcal{I}_{2}$ also converges.

\end{document}